# EVOLUTION LAW OF QUANTUM OBSERVABLES FROM CLASSICAL HAMILTONIAN IN NON-COMMUTATIVE PHASE SPACE

D. Dragoman – Univ. Bucharest, Physics Dept., P.O. Box MG-11, 077125 Bucharest, Romania, e-mail: danieladragoman@yahoo.com

ABSTRACT

The evolution equations of quantum observables are derived from the classical Hamiltonian equations of motion with the only additional assumption that the phase space is non-commutative. The demonstration of the quantum evolution laws is quite general; it does not rely on any assumption on the operator nature of $x$ and $p$ and is independent of the quantum mechanical formalism.

It is often believed that quantum mechanics is fundamentally different from classical mechanics, this difference being manifestly explicit in the quantum measurement theory, which implies the collapse of the wavefunction and which forbids the simultaneous precise determination of canonical conjugate variables. It is also thought that, although classical mechanics can be recovered as a $\hbar \to 0$ limit from quantum mechanics, the latter cannot be derived from the former. In particular, it is assumed that uncertainty relations such as $\Delta x \Delta p \geq \hbar/2$ are not general enough to derive quantum mechanics starting from classical mechanics. This point of view has been challenged in [1], where it was shown that the equations of motion of a quantum ensemble can be obtained from those of a classical ensemble from an exact uncertainty principle, provided that the classical ensemble undergoes random momentum fluctuations, which get rid of well-determined particle trajectory. The strength of these momentum fluctuations should be inversely correlated with the uncertainty in the particle position, which is not exactly known but is described by a position probability density. In another attempt to derive the Schrödinger equation from classical mechanics the classical point-like particle of mass $m$ was assumed to undergo a frictionless Brownian motion with a diffusion coefficient $\hbar/2m$, the form of which accounts for the lack of Brownian motion of macroscopic bodies [2]. A modified version of this model [3] assumes the existence of stochastic forces that produce a departure $\Delta E$ from the particle's classical energy, in energy conserving trajectories, which can persist for an average time $\Delta t \cong \hbar/(2\Delta E)$. These energy fluctuations originate in the energy exchange between the material particle and the vacuum (seen as an energy reservoir modeled in terms of virtual particles), and can allegedly explain phenomena thought to be purely quantum, such as the zero-point energy of oscillators, the stability of atoms, the slit diffraction, and the tunneling effect.

The common point of all these efforts to derive quantum mechanics from classical mechanics is the introduction of improbable, ad-hoc assumptions about the motion of the classical particle, which are not based on experimental results. However, the problem persists. It is still desirable to identify the element that leads to quantum mechanics from the classical theory of motion in order to understand better the quantum nature of many phenomena. Besides the uncertainty relations between non-commuting position and momentum observables, or energy and time, discussed above, the essence of quantum mechanics has been identified with the principle of superposition of states [4] or with commutation relations such as $[\hat{x}, \hat{p}] = i\hbar$ (see the references in [1]). According to the author's knowledge, however, no attempt has been made to derive quantum mechanics from these principles starting from classical mechanics. The aim of this paper is to show that it is possible to derive the evolution law of quantum observables from the classical Hamiltonian law of motion with the unique assumption that the position and momentum variables do not commute. We have been inspired in the demonstration method by the Feynman's proof of Maxwell equations [5].

Let us consider for simplicity a one-dimensional classical particle with position $x$ and momentum $p$, subject to a time-independent Hamiltonian $H$, i.e. to the classical equations of motion

$$\dot{p} = -\partial H / \partial x, \quad \dot{x} = \partial H / \partial p, \tag{1}$$

which moves in a non-commutative phase space, in which

$$[x, x] = 0, \quad [p, p] = 0, \quad [x, p] = xp - px = i\hbar. \tag{2}$$

We have not used in (2) the operator notation for *x* and *p*, since we do not make any assumption about their nature. The commutation relations (2) are suggested by experimental results.

The demonstration is much simplified if the Hamiltonian can be separated in a kinetic and a potential term, separation that is always possible in classical mechanics for a particle of mass *m* subject to a force that derives from a potential $V(x)$. Then, for $H = p^2/2m + V(x)$, we have

$$\dot{x} = p/m, \quad \dot{p} = -\partial V / \partial x, \tag{3}$$

and, after straightforward calculations, it follows from (2) that

$$[x, H] = [x, p^2/2m] = i\hbar p/2m + (p/2m)[x, p] = i\hbar p/m = i\hbar \partial H / \partial p. \tag{4}$$

We have thus obtained the quantum evolution law for position in a system with a time-independent Hamiltonian:

$$i\hbar \dot{x} = i\hbar \partial H / \partial p = [x, H]. \tag{5}$$

Note that in (5) *x* is the analog in a non-commutative phase space of the classical position. We refer to this *x* as an observable, since we assume it can be observed and measured as the classical *x* variable; after all we measure positions and angles (momentum components) in quantum mechanics. This definition of the *x* (and *p*) observable does correspond to the quantum definition since *x* and *p* satisfy the quantum commutation relation (2). However, no indication of any expectation value is to be found in (5); this equation is independent of the

Born's interpretation or of any formalism (Schrödinger, Heisenberg, etc.) of quantum mechanics.

The similar evolution law to (5) for the momentum observable can be found by employing the Jacobi identity

$$[H,[x,p]] + [x,[p,H]] + [p,[H,x]] = 0, \qquad (6)$$

in which the first term in the left-hand side vanishes since $[x, p]$ is a constant, and by noting that (1) and (2) imply

$$[\dot{x}, p] + [x, \dot{p}] = 0 = [\partial H / \partial p, p] + [x, -\partial H / \partial x]. \qquad (7)$$

By multiplying (7) with $-i\hbar$ and adding it to (6) one obtains

$$[x,([p,H] + i\hbar \partial H / \partial x)] + [p,([H,x] + i\hbar \partial H / \partial p)] = 0 \qquad (8)$$

from which, using (5), it follows that

$$i\hbar\dot{p} = -i\hbar \partial H / \partial x = [p, H]. \qquad (9)$$

In fact, from (8) it only results that (9) is a function of $x$, but this function can be shown to vanish if a certain form of the potential $V(x)$ is assumed.

Equations (5) and (9), which represent the evolution law under time-independent Hamiltonians for quantum observables, have been obtained from the classical Hamiltonian equations with the only additional assumption of phase space non-commutativity. In the newly obtained evolution laws no hypothesis have been made as to the nature of the $x$ and $p$, or about their significance as averages or expectation values. The only assumption was that $x$ and $p$ in the non-commutative phase space satisfy the same evolution equation as in classical mechanics.

The demonstration of the quantum evolution laws (5) and (9) can be generalized to any function of $x$ and $p$. Using (5) and (9), it is a straightforward task to show that

$$i\hbar d(x^n p^m)/dt = [x^n p^m, H],  \qquad (10)$$

and that a similar equation is satisfied by any function $F$ of $x$ and $p$, which does not explicitly depend on time:

$$\dot{F}(x,p) = [F,H]/i\hbar.  \qquad (11)$$

After this result is established, an explicit time dependence can be easily accounted for.

Since the classical evolution law for any function of $x$ and $p$ is known to be given by the Poisson bracket, (11) implies that in a classical non-commutative phase space the Poisson bracket $\{F,H\} = (\partial F/\partial x)(\partial H/\partial p) - (\partial H/\partial x)(\partial F/\partial p)$ has to be replaced by $[F,H]/i\hbar$, result that is known to hold in quantum mechanics. This result is obtained in quantum mechanics from the Schrödinger equation, the replacement rule $\{F,H\} \to [F,H]/i\hbar$ being observed to occur, but not demonstrated. In this paper a demonstration for this correspondence rule is provided in detail for the position and momentum observables, endowing the classical variables with a single new property: that of phase space non-commutativity.

Because $(xp - px)/2 = x \wedge p$ defines the outer product of $x$ and $p$, which equals the oriented area of the parallelogram with sides $x$ and $p$ [6], it follows that the essential feature of quantum mechanics is the replacement of the point-like classical particles with a "quantum" particle that, although evolving according to classical laws, is always localized in a phase space area $\hbar/2$. (Note that a similar change in phase space topology, from a point to an area equal to $\lambdabar/2$, with $\lambdabar = \lambda/2\pi$ and $\lambda$ the optical wavelength, characterizes the transition from ray optics to wave optics. In wave optics an operator $\hat{p} = -i\lambdabar\partial/\partial x$, analogous to the quantum mechanical operator $\hat{p} = -i\hbar\partial/\partial x$, can be introduced [7], which is canonically conjugate to the transverse position in the Hamiltonian sense if the longitudinal spatial coordinate plays the role of the time coordinate in quantum mechanics. Moreover, the quantum uncertainty relation is expressed in optics through the beam quality factor $Q$ [8]). This point of view is sustained by the demonstration that quantum wavefunction discontinuities propagate along classical trajectories in quantum mechanics, in a similar manner as electromagnetic field discontinuities propagate along rays in geometrical optics, the classical Hamilton-Jacobi equation corresponding to the eikonal equation in optics [9].

In conclusion, we can safely affirm that the commutation relation embodies the essence of quantum mechanics and allows the deduction of evolution laws for quantum observables (understood in this paper as power laws of momentum and position, which are measurable and thus observable quantities in a theory such as quantum mechanics that is based on the commutation relations (2)), irrespective of the quantum mechanical formalism. These laws are obtained without the need to introduce Hilbert spaces, quantum states or wavefunctions with debatable meaning. The transition from classical to quantum mechanics is similar to the transition from ray optics to wave optics, at least from the point of view of deriving evolution equations for observables, and therefore does not need a special mathematical apparatus. The Hilbert space and the operators and states defined on it only obscure the physical significance of quantum mechanics, which in essence is identical to the classical mechanics of extended particles in phase space. The employment of classical position and momentum variables to describe a quantum state, through the Wigner distribution function [10], in particular (see [11-16] for its properties and relations to the standard formulations of quantum mechanics), is perfectly suited for the description of quantum phenomena, including the measurement problem [16], as long as the Wigner distribution function is interpreted as a quasi-probability distribution in phase space. The negative value regions of the Wigner distribution function, which correspond to dark rays, do not only lead to positive probability values if averaged on the phase space area associated to the particle, but are necessary for the consistency of the theory since they are intimately related to wave-like phenomena such as diffraction and interference [17].

REFERENCES

[1]  M.J.W. Hall and M. Reginatto, "Schrödinger equation from an exact uncertainty principle", J. Phys. A **35**, 3289-3303, 2002

[2]  E. Nelson, "Derivation of the Schrödinger equation from Newtonian mechanics", Phys. Rev. **150**, 1079-1085, 1966

[3]  L. Fritsche and M. Haugk, "A new look at the derivation of the Schrödinger equation from Newtonian mechanics", Ann. Phys. (Leipzig) **12**, 371-403, 2003

[4]  P.A.M. Dirac, *The Principles of Quantum Mechanics*, 4th edition, Clarendon Press, Oxford, 1958

[5]  F.J. Dyson, "Feynman's proof of the Maxwell equations", Am. J. Phys. **58**, 209-211, 1990

[6]  D. Hestenes, *Space-Time Algebra*, Gordon and Breach Science Publishers, New York, 1966

[7]  D. Gloge and D. Marcuse, "Formal quantum theory of light rays", J. Opt. Soc. Am. **59**, 1629-1631, 1969

[8]  D. Dragoman, "The Wigner distribution function in optics and optoelectronics", Prog. Opt. **37**, 1-56, 1997

[9]  A. Luis, "Classical mechanics and the propagation of the discontinuities of the quantum wave function", Phys. Rev. A **67**, 024102, 2003

[10]  E. Wigner, "On the quantum correction for thermodynamic equilibrium", Phys. Rev. **40**, 749-759, 1932

[11]  M. Hillery, R.F. O'Connell, M.O. Scully, and E.P. Wigner, "Distribution functions in physics: Fundamentals", Phys. Rep. **106**, 121-164, 1984

[12]  H.-W. Lee, "Theory and application of the quantum phase-space distribution functions", Phys. Rep. **259**, 147-211, 1995

[13]  W.P. Schleich, *Quantum Optics in Phase Space*, Wiley-VCH, Berlin, 2001

[14]  Y.S. Kim and M.E. Noz, *Phase-Space Picture of Quantum Mechanics*, World Scientific, Singapore, 1991

[15]  D. Dragoman, "Phase space correspondence between classical optics and quantum mechanics", Prog. Opt. **43**, 433-496, 2002

[16]  D. Dragoman, "Phase space formulation of quantum mechanics. Insight into the measurement problem", Physica Scripta **72**, 290-295, 2005

[17]  E.C.G. Sudarshan, "Pencils of rays in wave optics", Phys. Lett. **73A**, 269-272, 1979